\documentclass[9pt,twocolumn,twoside]{opticajnl}
\journal{opticajournal} 

\setboolean{shortarticle}{true}

\usepackage{braket}
\usepackage{xcolor}
\usepackage{amsmath}
\newcommand{\RS}[1]{#1}   

\usepackage{lineno}

\title{Transverse Faraday Effect in Multimode Fibres}

\author[1]{Mohammed R. Sharif}
\author[1,*]{Mitchell A. Cox}

\affil[1]{School of Electrical and Information Engineering, The University of the Witwatersrand, Johannesburg 2050, South Africa}

\affil[*]{Mitchell.Cox@wits.ac.za}

\begin{abstract}
\RS{The weak guidance approximation suppresses the longitudinal field component \(E_z\), and with it an entire class of linear magneto-optic interactions. We derive and experimentally verify a transverse Faraday effect in step-index multimode fibres that couples \(E_y\) into \(E_z\). Using speckle decorrelation we isolate the linear Faraday signal from the quadratic Cotton--Mouton background, confirming four predictions: the effect vanishes without a strong index step, scales with launch angle, is completely polarisation-selective relative to the field, and tracks Verdet dispersion. Under 0.24~T we observe $\Delta\rho \approx 0.16$ for perpendicular input and a complete null for parallel input.}
\end{abstract}

\setboolean{displaycopyright}{false} 

\doi{} 
\begin{document}

\maketitle

Magneto-optic interactions in optical fibre underpin technologies from optical isolation to current sensing, and their theoretical description depends sensitively on which field components the guiding model retains. The weak guidance approximation simplifies fibre mode analysis by reducing the exact vector wave equation \cite{snyder1983optical}:
\begin{equation}
 \nabla^2\mathbf{E} - \nabla(\nabla\cdot \mathbf{E}) + k^2_0 \varepsilon_r \mathbf{E} = 0
 \label{eq:vector_wave_equation}
\end{equation}
\RS{Since \(\nabla\cdot (\varepsilon_r \mathbf{E}) = 0\), the divergence in the polarisation-coupling term \(-\nabla(\nabla\cdot\mathbf{E})\) expands to \(\nabla \cdot \mathbf{E} = - \mathbf{E} \cdot \nabla(\ln \varepsilon_r)\). }By assuming a vanishing index gradient (\(\nabla(\ln \varepsilon_r) \approx 0\)), the weak guidance approximation forces \(\nabla\cdot \mathbf{E} \approx 0\). This artificially suppresses the longitudinal electric field (\(E_z\)), masking critical non-reciprocal phenomena. Consequently, while longitudinal magnetic fields yield the robust linear Faraday effect, transverse fields are classically assumed to excite only the weak, reciprocal Cotton--Mouton effect \cite{cotton_mouton, Zvezdin1997}. However, a full vectorial treatment reveals that the core-cladding index discontinuity in step-index fibres forces \(E_z \neq 0\) to satisfy boundary conditions, activating a linear, non-reciprocal transverse \(E_y \leftrightarrow E_z\) coupling through the generic magneto-optic tensor \cite{Landau1984}. \RS{Engineered platforms synthesise these components intentionally \cite{Yan2022,Li2025}; step-index fibres possess this channel naturally, masked only by weak guidance.} In this Letter, we derive the theoretical framework for this transverse Faraday effect in multimode fibres, and we experimentally isolate the linear signal from the quadratic Cotton--Mouton background using speckle decorrelation.

\begin{figure}
 \centering
\includegraphics{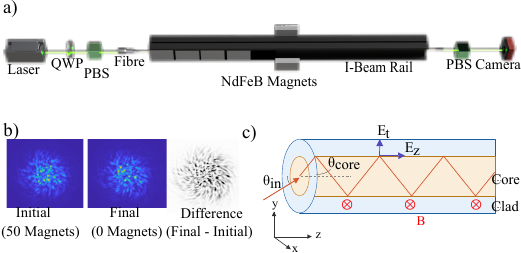}
 \caption{\RS{a) Experimental setup, with the fibre held in a custom I-beam rail for thermal and mechanical isolation. b) Speckle pattern evolution and residual. c) Field geometry: $\mathbf{B}=B_x\hat{x}$ is perpendicular to the $yz$-plane containing the ray. A ray launched at $\theta_{\rm in}$ propagates at internal angle $\theta_{\rm core}$, and tangential-field at each core-cladding reflection requires $E_z=|E|\sin\theta_{\rm core}$ alongside $E_t$.}}
 \rule{\linewidth}{0.5pt}
 \label{fig:Optical Setup}
 \vspace{-16pt} %
\end{figure}

To anchor this mechanism within established magneto-optics, we start from the most general linear constitutive relation for a lossless dielectric, \(D_i = \varepsilon_{ij}E_j\). Energy conservation in a lossless medium requires $\varepsilon_{ij}$ to be Hermitian (\(\varepsilon_{ij}=\varepsilon_{ji}^*\)) \cite{Landau1984}, so its real part is symmetric and its imaginary part is antisymmetric. Writing the antisymmetric part of \(\varepsilon_{ij}\) as \(b_{ij}\), any real antisymmetric \(3\times3\) matrix is equivalent to a pseudo-vector \(g\) via \(b_{ij}=\varepsilon_{ijk}g_k\). Consequently, the most general lossless permittivity tensor for an otherwise isotropic medium of index \(n\) takes the form:
\begin{equation}
\varepsilon_{ij} = \varepsilon_0 n^2\left(\delta_{ij} + i\varepsilon_{ijk}g_k\right),
\label{eq: Gyrotropic Tensor}
\end{equation}
where \(g\) is the gyration vector mediating the magneto-optic perturbation \cite{Zvezdin1997}. Contracting Eq.~\ref{eq: Gyrotropic Tensor} with \(E_j\) and using \(\varepsilon_{ijk}g_kE_j=(E\times g)_i\) yields the electric displacement field:
\begin{equation}
 D= \varepsilon_0 n^2[E+i(E \times g)]
\label{eq: E field Displacement}
\end{equation}
In the standard longitudinal Faraday effect, \(\mathbf{B} = B_z\hat{z}\) orients the gyration vector along the propagation axis, \(g=g_z\hat z\), coupling the transverse components (\(E_x \leftrightarrow E_y\)) \cite{Saleh2019}. This case anchors the absolute scale of \(g\) to the empirical Verdet constant. For \(\mathbf{B} = B_x \hat{x}\), Onsager reciprocity (\(\varepsilon_{ij}(\mathbf B) = \varepsilon_{ji}(-\mathbf B)\)) requires \(g\) to be an odd function of \(\mathbf B\). To leading order, \(g\) is linear in \(\mathbf B\) and aligns with it, yielding \(g = g_x\hat x\). Evaluating the cross product:
\begin{equation}
E \times g = E \times{g_x}\hat{x} = {g_x} (0, E_z, -E_y)^\top
\label{eq: cross product}
\end{equation}
This perturbation acts as a linear operator \(\Delta\varepsilon_{TF}\cdot \mathbf{E} \equiv i n^2 (\mathbf{E}\times \mathbf{g})\)~\cite{Zvezdin1997}. Its \(y\)-component returns \(E_z\) and its \(z\)-component returns \(-E_y\), giving:
\begin{equation}
\Delta \varepsilon_{TF} = 
\begin{pmatrix}
{xx} & {xy} & {xz} \\ 
{yx} & {yy} & {yz} \\ 
{zx} & {zy} & {zz}
\end{pmatrix}
= i n^2 g_x 
\begin{pmatrix}
0 & 0 & 0 \\ 
0 & 0 & 1 \\ 
0 & -1 & 0
\end{pmatrix}
\label{eq: TF}
\end{equation}
The explicit factor of \(i\) ensures that this skew-symmetric matrix in Eq.~\ref{eq: TF} is Hermitian (\(\Delta\varepsilon_{TF}^\dagger = \Delta\varepsilon_{TF}\)) for a lossless medium. The vanishing \(x\)-row confines the perturbation exclusively to coupling transverse \(y\)-components with longitudinal \(z\)-components. This coupling depends strictly on launch geometry: translating the input beam along the \(y\)-axis confines excited rays primarily to the \(yz\)-plane \RS{(Fig.~\ref{fig:Optical Setup}(c))}. For this trajectory, an \(x\)-polarised input (parallel to \(B\)) is a pure Transverse Electric (TE) wave (\(E_z = 0\)), nullifying the coupling integral (Eq.~\ref{eq: Coupling Coefficient}). Conversely, a \(y\)-polarised input forms a Transverse Magnetic (TM) wave; the non-zero \(E_z\) component required by boundary conditions then strongly activates the magneto-optic coupling.

\RS{The transverse coupling strength depends on the dimensionless gyration vector \(g_x\), which we map to the macroscopic Verdet constant \(V_d\) via the longitudinal geometry. There, \(g = g_z\hat{z}\) induces circular birefringence between eigenmodes}:
\begin{equation}
n_\pm \approx n\left(1 \pm \frac{g_z}{2}\right),
\label{eq: circular birefringence}
\end{equation}
After a length \(L\), the accumulated linear polarisation rotation is:
\begin{equation}
\theta_F = \frac{k_0}{2}(n_+-n_-)L = \frac{\pi}{\lambda}\,n\,g_z\,L.
\label{eq: longitudinal rotation}
\end{equation}
Equating this to the empirical definition \(\theta_F = V_d B_z L\) and exploiting the magnitude's isotropy (\(|g|\) depends only on \(|\mathbf{B}|\)), the transverse parameter becomes:
\begin{equation}
g_x = \frac{\lambda V_d}{\pi n}\,B_x \equiv f B_x, \qquad f = \frac{\lambda V_d}{\pi n}.
\label{eq: gx verdet}
\end{equation}
For fused silica, the dispersion of the Verdet constant yields approximately 5.2 rad T\(^{-1}\) m\(^{-1}\) at 532 nm and 3.7 rad T\(^{-1}\) m\(^{-1}\) at 633 nm \cite{Verdet}.

Eq.~\ref{eq: TF} is the specific correction that the material-response term $k_0^2\varepsilon_r\mathbf{E}$ acquires under a transverse magnetic field. Writing $\varepsilon_{r,ij} = n^2\delta_{ij} + \Delta\varepsilon_{TF,ij}$ and substituting into Eq.~\ref{eq:vector_wave_equation}, the term splits into an unperturbed piece $k_0^2 n^2 \mathbf{E}$, \RS{generating }the ordinary weakly-guided modes, and a first-order perturbation $k_0^2\Delta\varepsilon_{TF}\cdot\mathbf{E}$. \RS{F}irst-order coupled-mode theory \cite{snyder1983optical} projects this onto the unperturbed basis $\{\mathbf{E}_m\}$:
\begin{equation}
\kappa_{mn} = \frac{\omega\varepsilon_0}{4P} \iint_A \mathbf{E}_m^* \cdot \Delta\varepsilon \cdot \mathbf{E}_n \, dA
\label{eq: CMT general}
\end{equation}
Substituting Eq.~\ref{eq: TF}, the matrix product evaluates to:
\begin{equation}
\Delta\varepsilon_{TF}\cdot\mathbf{E}_n = in^2g_x\left(0,\ E_{z,n},\ -E_{y,n}\right)^\top
\end{equation}
Taking the dot product with \({E}_m^{*}\) then gives \({E}_m^{*} \cdot (\Delta \varepsilon_{TF}) \cdot E_n = i n^2 g_x ({E}_{y,m}^{*}{E}_{z,n}- {E}_{z,m}^{*}{E}_{y,n})\). Substituting back:
\begin{equation}
\kappa_{mn} = \frac{i \omega \varepsilon_0 n^2 g_x}{4P} \iint_A \left( E_{y,m}^* E_{z,n} - E_{z,m}^* E_{y,n} \right) dx\,dy,
\label{eq: Coupling Coefficient}
\end{equation}
This overlap vanishes when $E_z \approx 0$, explaining why the effect is invisible under the weak guidance approximation. The divergence condition gives $E_z \propto -\frac{i}{k_0 n}\nabla_t\cdot \mathbf E_t$ \cite{snyder1983optical}\RS{, so for} a $y$-polarised mode\RS{} $E_{z,n}\propto \partial E_{y,n}/\partial y$. Fibre eigenmodes carry an azimuthal harmonic $\propto F_l(r)\cos(l\varphi)$\RS{, and $\partial/\partial y$} raises and lowers the azimuthal order, \RS{so }$E_{z,n}$ carries only harmonics $l_n\pm1$\RS{. The overlap therefore} vanishes by angular orthogonality unless $l_m = l_n \pm 1$\RS{: }coupling survives only between cross-polarised modes of adjacent azimuthal order\RS{, and same-order} co-polarised self-coupling integrates to zero.

To quantify the surviving coupling, consider a polarised ray propagating at internal angle \(\theta_{core}\) \RS{(Fig.~\ref{fig:Optical Setup}(c))}. Since \(E\) must lie perpendicular to the propagation direction:
\begin{equation}
|E_z| = |E|\sin\theta_{\mathrm{core}}, \qquad |E_t| = |E|\cos\theta_{\mathrm{core}}.
\label{eq:field_components}
\end{equation}
So \(\eta =|E_z|/|E_t| \mathrel{\RS{\approx}} \tan \theta_{core}\). Applying Snell's law at the input face (\(\sin \theta_{in}=n_{core} \sin\theta_{core}\)) \cite{snyder1983optical}, $\cos\theta_{core}=\sqrt{1-\sin^2\theta_{core}}=\sqrt{n_{core}^2-\sin^2\theta_{in}}/n_{core}$, so that
\begin{equation}
\eta \equiv \frac{|E_z|}{|E_t|} \approx \tan\theta_{\text{core}} = \frac{\sin\theta_{\text{in}}}{\sqrt{n_{\text{core}}^2 - \sin^2\theta_{\text{in}}}}
\label{eq: Vector Tilt}
\end{equation}
\RS{Solving the exact vector characteristic equation, with no weak-guidance approximation, for every guided mode family of both test fibres at both wavelengths gives a modal longitudinal fraction $\eta_{\rm modal}=(\int|E_z|^2dA/\int|E_t|^2dA)^{1/2}$, evaluated in the unperturbed basis, proportional to Eq.~\ref{eq: Vector Tilt} throughout, with $\eta_{\rm modal}/\eta_{\rm ray} = 0.66 \pm 0.01$ across all four cases. The ray estimate therefore reproduces the scaling while overestimating the magnitude by \(\approx1.5\times\).}

\RS{E}valuating Eq.~\ref{eq: Coupling Coefficient} over hundreds of mode pairs is intractable, so we replace the mode-specific $E_{z,n}$ by a local field ratio $E_z(x,y) \approx \eta E_t(x,y)$ fixed by ray angle rather than mode index. \RS{Because coupling is restricted to a dense set of adjacent-order cross terms, summing over the coupled set is equivalent, at this level of estimate}, to a single overlap integral of the local transverse intensity, via the closure
\begin{equation}
\sum_n E_{y,m}^{*}E_{z,n} \;\to\; \eta\sum_n E_{y,m}^{*}E_{t,n} = \eta\,|E_t(x,y)|^2
\label{eq: local closure}
\end{equation}
taken over the coupled-mode set. Substituting this relation into Eq.~\ref{eq: Coupling Coefficient}, the effective coupling rate before normalisation reads:
\begin{equation}
\kappa_{\text{eff}} \approx \frac{\omega\varepsilon_0 n^2 g_x \eta}{4P}\iint_A |E_{t}|^2\,dA.
\label{eq: kappa pre-norm}
\end{equation}
The modal power normalisation, given by the time-averaged Poynting flux of a weakly-guided mode \cite{snyder1983optical} gives:
\begin{equation}
P \approx \frac{1}{2} c \varepsilon_0 n \iint_A |E_t|^2\, dA \;\;\Rightarrow\;\; \iint_A|E_t|^2\,dA = \frac{2P}{c\varepsilon_0 n},
\label{eq: Power Normalisation}
\end{equation}
the spatial integral cancels exactly against $P$ in Eq.~\ref{eq: kappa pre-norm}, leaving a mode-independent effective coupling rate
\begin{equation}
\kappa_{\text{eff}} = \frac{\omega\varepsilon_0 n^2 g_x \eta}{4P}\cdot\frac{2P}{c\varepsilon_0 n} = \frac{\omega n g_x \eta}{2c} = \frac{1}{2}k_0\,n\,g_x\,\eta,
\label{eq: Effective Coupling Rate}
\end{equation}
using $k_0=\omega/c$. Substituting the Verdet relation $g_x = (\lambda V_d/\pi n)B$ from Eq.~\ref{eq: gx verdet} and $k_0 = 2\pi/\lambda$:
\begin{equation}
\kappa_{\text{eff}} = \frac{1}{2}\cdot\frac{2\pi}{\lambda}\cdot n\cdot\frac{\lambda V_d}{\pi n}B\cdot\eta = V_d B \eta,
\label{eq: kappa final}
\end{equation}
where every factor of $n$ and $\lambda$ cancels exactly. \RS{Over a fully coherent interaction length \(L\), this yields the transverse Faraday rotation angle}
\begin{equation}
\Theta_{\mathrm{eff}}(\theta_{\mathrm{in}}) = \kappa_{\rm eff}L = V_d B L \cdot \eta(\theta_{\mathrm{in}}) = V_d B L \cdot \frac{\sin\theta_{\mathrm{in}}}{\sqrt{n_{\mathrm{core}}^2 - \sin^2\theta_{\mathrm{in}}}}.
\label{eq: Tunable Model}
\end{equation}

\begin{figure*}[t]
 \centering
 \vspace{-8pt} 
 \includegraphics[width=0.90\linewidth]{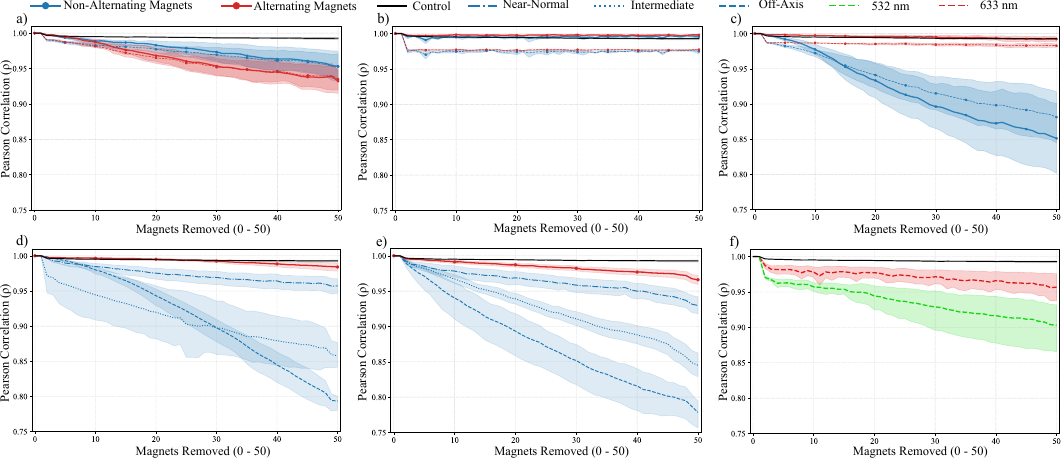}
 \caption{Isolation of Transverse Faraday effect using alternating and non-alternating magnetic fields. Panels (a) to (c) demonstrate polarisation selectivity in 105~$\mu$m fibre for \RS{right-handed circular ($\ket{R}$), parallel ($\ket{\parallel}$) and perpendicular ($\ket{\perp}$) inputs, defined relative to $\mathbf{B}$}. Panels (d) and (e) compare the angular scaling for 50~$\mu$m and 105~$\mu$m cores. Panel (f) illustrates the wavelength dependence at 532~nm and 633~nm. }
 \rule{\linewidth}{0.5pt}
 \vspace{-16pt} %
 \label{fig: All Results}
\end{figure*}

\RS{Since \(\Theta_{\mathrm{eff}} = V_d B (\eta L)\) has the form \(\theta_F = V_d B L_{\mathrm{eff}}\), the product \(\eta L\) is the equivalent longitudinal interaction length producing the same rotation; \(\eta\) is an amplitude ratio, so \(\eta L\) is amplitude-weighted rather than a physical propagation distance. Launch was set by transverse mirror displacement rather than by calibrated angle, so the three configurations are ordinal rather than resolved. Evaluating \(\eta L\) across the guided range for illustration gives \(\eta L \approx 23\), 82 and 135~mm at \(\theta_{\mathrm{in}} = 2^\circ\), \(7^\circ\) and the NA limit \(11.5^\circ\) (\(\mathrm{NA}=0.20\), \(n_{\mathrm{core}}=1.491\), \(L=1.00\)~m). The ordering of \(\Delta\rho\) in Fig.~\ref{fig: All Results}(d--e) follows that of \(\eta\).}

\RS{Because the selection rule mandates coupling between cross-polarised modes of adjacent azimuthal order, each coupled pair \((m,n)\) accumulates amplitude subject to a phase mismatch \(\Delta\beta_{mn}\),}
\begin{equation}
a_{mn} = \kappa_{\rm eff}\int_0^L b(z)\,e^{i\Delta\beta_{mn}z}\,dz , \qquad b(z) = B_x(z)/B ,
\label{eq: pair amplitude}
\end{equation}
\RS{In the non-alternating configuration \(b(z)=1\), giving \(|a_{mn}|^2 = \kappa_{\rm eff}^2\,4\sin^2(\Delta\beta_{mn}L/2)/\Delta\beta_{mn}^2\), sharply peaked at \(\Delta\beta \to 0\) and falling as \(\Delta\beta^{-2}\) elsewhere. Summing over the coupled set with pair density \(p(\Delta\beta)\), and using \(4\sin^2(\Delta\beta L/2)/\Delta\beta^2 \to 2\pi L\,\delta(\Delta\beta)\) for \(\Delta\beta_{\rm typ} L \gg 1\):}
\begin{equation}
\langle \Theta^2 \rangle = \kappa_{\mathrm{eff}}^{2}\,\ell_c L \propto V_d^{2} B^{2} \eta^{2} \ell_c L , \qquad \ell_c \equiv 2\pi p(0)
\label{eq: Phase Variance}
\end{equation}
\RS{The dependence on \(L\) is linear rather than quadratic because the phase-matched bandwidth narrows as \(1/L\). With \(L_b \sim 1\)~mm, \(\ell_c \ll L\) and the coherent result of Eq.~\ref{eq: Tunable Model} is suppressed accordingly.} \RS{We treat} the highly multimode output as a fully developed circular complex Gaussian speckle field\RS{; applying} the Siegert relation \cite{Goodman2007}, the intensity correlation decays exponentially with the accumulated phase variance:
\begin{equation}
 \rho(B, L, \theta_{\rm in}) \approx \exp\!\left(-\alpha'(\theta_{\rm in})\, B^2 L\right), \quad \alpha' \propto V_d^2\, \eta^2(\theta_{\rm in})\RS{\ell_c}.
\label{eq: Predicted PCC}
\end{equation}
For \RS{\(\langle\Theta^2\rangle \ll 1\)}, \RS{expanding }Eq.~\ref{eq: Predicted PCC} \RS{gives} $\rho \approx 1 - \RS{\langle\Theta^2\rangle} \propto B^2$, the quadratic field dependence \RS{characteristic of} linear magneto-optic accumulation. \RS{T}he Cotton--Mouton phase shift is \RS{itself quadratic in $B$, so its} intensity decorrelation scales as $B^4$\RS{ and is vanishingly} small at these field strengths, justifying its treatment as a stable, flat baseline. \RS{The coherent single-pair estimate \(\Theta_{\rm eff} = V_d B L \eta \approx 0.17\)~rad gives \(\langle\Theta^2\rangle \approx 3\times10^{-2}\); the scalar closure of Eq.~\ref{eq: local closure} collapses \(\sim\!10^4\) coupled pairs onto a single overlap and does not fix the absolute prefactor of \(\alpha'\). The predictions tested below are accordingly scaling relations.} The framework yields four testable predictions: (1) the effect vanishes without a strong index step ($E_z \approx 0$); (2) the signal scales monotonically with $\theta_{\rm in}$; (3) parallel input ($\ket{\parallel}$) nulls completely while perpendicular ($\ket{\perp}$) maximises the $E_z$-mediated coupling; and (4) the decorrelation rate tracks $V_d(\lambda)$, distinguishing it from the quadratic Cotton--Mouton baseline.

We track the phase-driven spatial mode mixing caused by the magnetic perturbation by monitoring $\rho$ of the output intensity speckle pattern relative to an initial reference state \cite{Goodman2007, Ploschner2015}. \RS{Because the coupled pairs are strongly phase-mismatched (Eq.~\ref{eq: Phase Variance}), the accumulated rotation has vanishing mean while its variance grows with length. A polarimetric measurement returns the mode-summed Stokes vector and is therefore sensitive to precisely the first moment that averages to zero here; with \(10^3\text{--}10^4\) modes it is in any case strongly depolarised. The speckle pattern retains the spatial structure of the mode superposition, so a redistribution of relative phase registers directly in the intensity. The output was nonetheless polarisation-resolved: a PBS before the camera separated the components parallel and perpendicular to \(\mathbf{B}\), which is how Fig.~\ref{fig: All Results}(a--c) was obtained.}

To isolate the linear transverse Faraday effect from the quadratic Cotton--Mouton effect we used two magnet configurations. An alternating arrangement, w\RS{ith successive pairs of} reversed polarity, unwinds and cancels any linear phase accumulation \RS{and so isolates the quadratic effect}. \RS{A} non-alternating configuration (\RS{$\int B_x\,dz \neq 0$}) allows the linear coupling to accumulate along the fibre.

A 532~nm laser passed through a quarter-wave plate (QWP) and polarising beam splitter (PBS) \RS{selecting} one of four input states relative to \RS{$\mathbf{B}$}: parallel ($\ket{\parallel}$), perpendicular ($\ket{\perp}$), right-handed ($\ket{R}$) \RS{and} left-handed ($\ket{L}$) \RS{circular}. Launch angle was \RS{set by translating} a steering mirror before the QWP along a transverse \RS{($y$)} translation stage\RS{, giving} near-normal, intermediate and highly off-axis incidence. 

Measurements were conducted at 532~nm and 633~nm on 1.0~m lengths of four fibre types: a few-mode step-index fibre (SMF-28) and a 50~$\mu$m graded-index fibre serving as controls, alongside 50~$\mu$m and 105~$\mu$m step-index multimode fibres ($\mathrm{NA} = 0.20$) serving as the test waveguides where boundary reflections generate finite $E_z$. The output speckle pattern was imaged directly onto a camera, with each fibre secured in the custom 3D-printed I-beam rail of Fig.~\ref{fig:Optical Setup}. N32 neodymium permanent magnets were arranged along the rail in opposing pairs (3.5~mm transverse gap, \RS{20.4}~mm longitudinal pitch), subjecting the fibre to a transverse magnetic field ($\mathbf{B} = B_x \hat{x}$) of $B = 0.24$~T (\RS{finite-element estimate}).

\RS{M}easurements proceeded by sequentially removing magnet pairs from a populated rail of 50, recording a speckle image and computing the Pearson correlation against the initial 50-pair reference after each removal, over $n = 10$ trials with the fibre and launch optics left undisturbed to maintain a constant launch angle.

\RS{Baseline stability was confirmed via \(n=300\) control runs, duration-matched (\(\approx 90\)~s) and interleaved with active runs; \(\rho \approx 0.99\) across all fibres}.

\RS{Throughout, \(\Delta\rho \equiv \rho_{\mathrm{ref}} - \rho\) after 50 removals, where \(\rho_{\mathrm{ref}}\) is the alternating-configuration trace where one was recorded (Fig.~\ref{fig: All Results}(a--e)) and the unperturbed control run otherwise (Fig.~\ref{fig: All Results}(f)). An environmental origin is excluded because: (1) environmental noise is polarisation-blind, whereas the response exhibits strict state-dependent extrema (\(\Delta\rho \approx 0\) parallel versus \(\Delta\rho \approx 0.16\) perpendicular, Fig.~\ref{fig: All Results}(b--c)); (2) the alternating configuration undergoes identical handling yet maintains a flat baseline; and (3) decorrelation scales systematically with launch angle, core size, and \(V_{d}(\lambda)\).}

\RS{Prediction 1 was tested on the few-mode step-index fibre (SMF-28 at 532~nm) and the 50~$\mu$m GRIN fibre: after 50 removals the two configurations differ by \(\Delta\rho = +0.008 \pm 0.014\) and \(-0.044 \pm 0.066\) respectively, both consistent with zero over \(n=10\) trials and the latter of opposite sign to the transverse Faraday signature. Neither meets the geometric prerequisite: the graded-index profile lacks an abrupt boundary, and the SMF-28 supports too few modes at 532~nm (\(V \approx 6.8\)) for speckle correlation to resolve a redistribution of modal phase, so subsequent measurements focus on the step-index multimode fibres, where the effect is active.}

\RS{Prediction 3 is tested in Fig.~\ref{fig: All Results}(a--c) using the 105~$\mu$m fibre at intermediate incidence. Parallel input (\(\ket{\parallel}\)) completely suppresses the effect: the alternating and non-alternating trajectories overlap, showing no measurable decorrelation (\(\Delta\rho \approx 0\)). Perpendicular input (\(\ket{\perp}\)) yields the maximum decorrelation (\(\Delta\rho \approx 0.16\), \(\rho \approx 0.84\)). The circular states (\(\ket{R}\), \(\ket{L}\)) give identical intermediate responses (\(\rho \approx 0.95\)--\(0.96\)), and the observed hierarchy (\(\ket{\parallel} < \text{circular} < \ket{\perp}\)) matches the zero structure of the \(\Delta\varepsilon_{\text{TF}}\) tensor (Eq.~\ref{eq: TF}).}

\RS{Prediction 2 requires \(\Delta\rho\) to scale monotonically with the vector tilt (Eq.~\ref{eq: Tunable Model}). Figure~\ref{fig: All Results}(d--e) shows \(\rho\) versus magnets removed for the 50 and 105~$\mu$m step-index fibres at three launch angles.} In the alternating configuration, all three angles produce overlapping curves that remain stable at $\rho \approx 0.98$ and $\rho \approx 0.97$ for the 50~$\mu$m and 105~$\mu$m fibres, respectively. The complete suppression of the linear Faraday signal in the alternating configuration isolates the quadratic Cotton--Mouton baseline: \RS{the alternating field has a zero-mean spatial distribution, displacing its Fourier weight from \(\Delta\beta = 0\) to \(\pm\pi/p\). Coupling would then require a beat length \(2p = 40.8\)~mm, which no adjacent-group pair possesses, so the DC spatial component driving Eq.~\ref{eq: Phase Variance} is removed.}

\RS{Conversely, the non-alternating configuration decreases monotonically. For the 50~$\mu$m fibre, near-normal incidence gives \(\Delta\rho \approx 0.025\) (\(\rho \approx 0.955\)), rising to \(\Delta\rho \approx 0.125\) at intermediate incidence and \(\Delta\rho \approx 0.18\) (\(\rho \approx 0.80\)) at the steepest launch. The 105~$\mu$m fibre follows the same trend more strongly, reaching \(\Delta\rho \approx 0.19\) (\(\rho \approx 0.78\)) against the same \(\rho \approx 0.97\) alternating baseline. At the steepest launch the two cores agree to \(\approx 5\%\). Plotted against magnet pairs remaining, \(\ln\rho\) is linear for both cores, consistent with the \(\exp(-\alpha' B^2 L)\) form of Eq.~\ref{eq: Predicted PCC} and with the linear \(L\)-dependence of Eq.~\ref{eq: Phase Variance}.}

Finally, we \RS{compare the} 105~$\mu$m fibre under 532~nm and 633~nm illumination at the same near-normal launch angle (Fig.~\ref{fig: All Results}(f)). The 532~nm source \RS{decorrelates further,} reaching $\rho \approx 0.90$ \RS{against} $\rho \approx 0.95$ \RS{at} 633~nm\RS{, giving \(\Delta\rho_{532} \approx 0.09\) against \(\Delta\rho_{633} \approx 0.04\). The measured ratio \(\approx 2.2\) is consistent with the Verdet scaling \((V_{532}/V_{633})^2 \approx 2.0\), the small excess following from the weak wavelength dependence of \(\ell_c\), confirming Prediction 4.}

In conclusion, we have experimentally isolated the transverse Faraday effect in step-index multimode fibres. Under 0.24~T, we observe $\Delta \rho \approx 0.16$ in the 105~$\mu$m fibre for perpendicular input and a complete null for parallel input, with all four predictions confirmed. This reveals an $E_y \leftrightarrow E_z$ coupling channel rendered invisible by the weak guidance approximation.

\begin{backmatter}
\bmsection{Funding}
Council for Scientific and Industrial Research 
\bmsection{Disclosures}
The authors declare no conflicts of interest.
\end{backmatter}
\bibliography{sample}


\end{document}